# Photoacoustic characteristics of carbon-based infrared absorbers

Jussi Rossi[1], Juho Uotila[2], Sucheta Sharma[3], Toni Laurila[3], Roland Teissier[4], Alexei Baranov[4], Erkki Ikonen[3,5] and Markku Vainio[1,6]

1 Photonics Laboratory, Physics Unit, Tampere University, Tampere, Finland
2 Patria Aviation Oy, Tampere, Finland
3 Metrology Research Institute, Aalto University, Espoo, Finland
4 IES, University of Montpellier, CNRS, 34095 Montpellier, France
5 VTT MIKES, Espoo, Finland
6 Department of Chemistry, University of Helsinki, Helsinki, Finland
Contact: jussi.rossi@tuni.fi or markku.vainio@helsinki.fi

**Abstract**

We present an experimental comparison of photoacoustic responsivities of common highly absorbing carbon-based materials. The comparison was carried out with parameters relevant for photoacoustic power detectors and Fourier-transform infrared (FTIR) spectroscopy: we covered a broad wavelength range from the visible red to far infrared (633 nm to 25 µm) and the regime of low acoustic frequencies (< 1 kHz). The investigated materials include a candle soot-based coating, a black paint coating and two different carbon nanotube coatings. Of these, the low-cost soot absorber produced clearly the highest photoacoustic response over the entire measurement range.

**Introduction**

In addition to its many applications in spectroscopy [1-4] and imaging [5,6], the photoacoustic (PA) effect is useful for electromagnetic power detection due to its wavelength independency and high detection sensitivity. In a typical photoacoustic optical power detector, the incident radiation is first modulated by a chopper and then directed through a window to a PA cell. The cell contains an optical absorber to generate an acoustic wave at the chopping frequency. It is filled with gas that carries the acoustic signal to a sensitive microphone, whose output is proportional to the optical power incident on the detector. An example of a commonly used PA detector is Golay cell, in which the signal is recorded by optical readout of a thin reflective membrane that stretches due to the acoustic wave [7]. Although the photoacoustic detection principle works at practically any wavelength, it is mainly used in the infrared and terahertz (THz) regions, where it is one of the most sensitive power detection methods available [8-11].

An essential component of the PA power detector is the absorber. An ideal (hyperblack) absorber would have a flat and perfect (100 %) absorbance at all wavelengths [12]. The broad and uniform spectral responsivity is important not only for general-purpose power detectors but also from the metrological point of view: Traceable power measurements in the infrared and THz regions benefit from the possibility of transferring the calibration to the visible wavelength region, where a more accurate responsivity scale is available [13]. Another example of an application that requires an optically broadband absorber is photoacoustic Fourier Transform Infrared (FTIR) spectroscopy, where highly absorptive carbon reference materials are used to normalize FTIR spectra of unknown samples [14]. In other words, the spectrum of an unknown sample is measured and divided by the FTIR spectrum of the reference absorber. This procedure removes the spectral dependence of the FTIR instrument if the reference absorber has a uniform and/or well-characterized spectral responsivity throughout the whole measurement range [15].

While the emissivities of different black materials have been extensively investigated [16-18], information about their photoacoustic properties is difficult to find in the literature. Detailed studies of PA efficiencies of different carbon-based absorbers in the visible wavelengths have been published in view of ultrasound generation for medical applications [19,20]. However, as far as we know, comparisons of photoacoustic properties of absorber materials for conditions relevant in optical power detection and FTIR measurements are yet to be reported. These applications require information about photoacoustic performance of different materials at low acoustic frequencies (from approximately 10 Hz to 100 Hz) and over a wide range of wavelengths, particularly in the infrared part of the spectrum.

In this paper, we report an experimental comparison of the photoacoustic responsivities of different carbon-based absorbers over a wide wavelength range, from the visible red to far infrared (25 µm). The results of this research benefit the development of next-generation photoacoustic power detectors, traceable long-wavelength power measurements, as well as photoacoustic FTIR spectroscopy. We have also studied the dependence of signal strength on modulation (chopping) frequency and acoustic carrier gas for our experimental setup, which is based on a silicon cantilever microphone. The cantilever-enhanced photoacoustic method has already led to some of the best detection sensitivities in photoacoustic trace-gas spectroscopy [3,4,21], and it has the potential to significantly advance optical power detector development as well.

**Photoacoustic instrument and absorber materials**

The experiments described in this paper were carried out using a commercially available photoacoustic detector (PA301, Gasera Ltd). The detector is originally designed as an accessory for FTIR analysis of solid and liquid samples. The incoming optical power is first modulated with a chopper and then collected with a gold-coated ellipsoid mirror. The mirror guides the light beam through the KBr window of the photoacoustic cell and to the center of the sample, which in our case is the absorber under study. The instrument works over a wide wavelength range, from approximately 0.6 to 25 µm. The long-wavelength side is limited by the transparency of the KBr window, while the short-wavelength limit is determined by the gold-coated ellipsoid mirror.

The acoustic signal generated at the absorber is recorded with a cantilever microphone fabricated from silicon [22]. The cantilever is designed for the detection of low acoustic frequencies (< 1 kHz), where it shows a large linear dynamic range and high detection sensitivity [23]. The photoacoustic cell is filled with an acoustic carrier gas, which in our measurements is typically air, $N_2$ or He. The gas pressure is the same as the ambient pressure. The absorber under study is placed in a 10-mm diameter sample cup made of aluminum. The sample cup is located behind the window of the photoacoustic cell.

**Absorbers**

The photoacoustic comparison was carried out with a set of black absorbers that potentially work both in the visible and infrared parts of the electromagnetic spectrum. As relevant prior information of the photoacoustic properties of different absorbers is scarce, we selected the samples mainly based on the previous studies concerning the emissivity, reflectivity and ultrasonic photoacoustic conversion efficiency of various black materials [16-20,24-26]. The selected absorbers include two commercially available carbon nanotube (CNT) surfaces, candle soot and a commercially available ultrablack paint (Nextel), all of which are briefly described below. Each absorber was fabricated on an aluminum substrate. The absorber thicknesses were determined by scanning electron microscope (SEM) measurements, see Fig. 1.

Carbon nanotube coatings were chosen for the comparison because they are known to have low reflectances from the visible to the infrared region. In addition, vertically aligned CNT arrays have provided excellent performance in thermopile and pyroelectric thermal detectors [12,24,27,28]. Our CNT absorbers were ordered from Surrey Nanosystems Ltd. Two different samples were selected: Spray-applied coating of randomly oriented CNTs (S-VIS, Fig. 1a-b) and a surface that consists of vertically aligned CNT arrays (Vantablack, Fig. 1c-d). The thicknesses of the absorbing CNT layers are approximately 80 µm and 200 to 400 µm for the Vantablack and S-VIS samples, respectively. The manufacturer has fabricated the absorbing layers on aluminum substrate, which were cut to fit in the 10-mm sample cup of our PA detector. Extreme care was taken to not to touch the fragile sample surfaces.

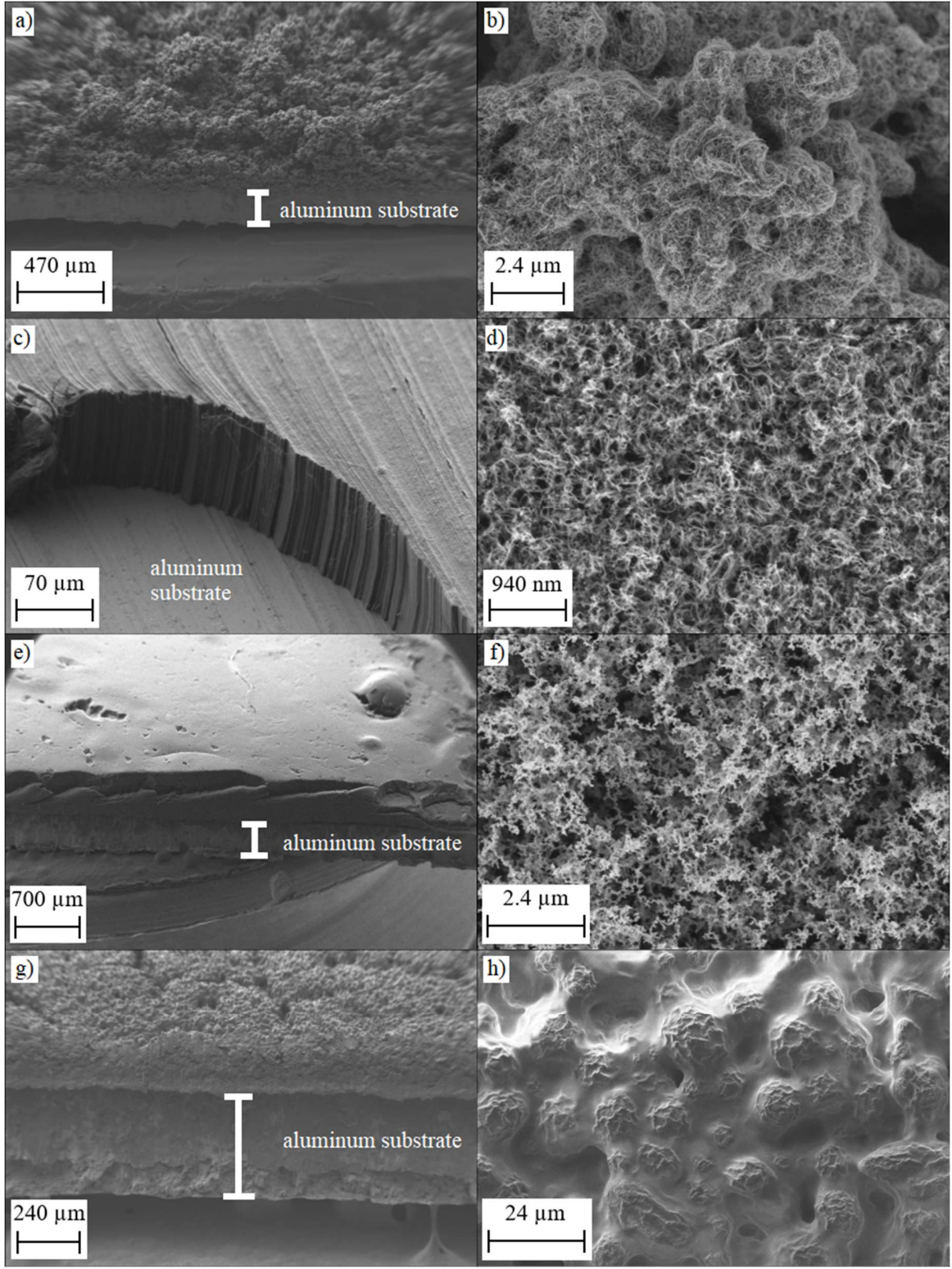

**Figure 1.** Side view (left row) and top view (right row) SEM pictures of the cut samples. a) – b) S-VIS, c) – d) Vantablack, e) – f) candle soot and g) – h) Nextel-coating.

Among other applications, candle soot coatings have been used in efficient pyroelectric energy conversion [16,17] and ultrasound generation [19]. Although the high specific heat capacity of carbon-based soot and paint coatings is not necessarily ideal for other thermal detectors [25], they are potentially useful materials in photoacoustic detection. Our soot absorbers (Fig. 1e-f) were fabricated directly on the bottom of the aluminum sample cup using a paraffin wax candle flame (Fig. 2a). The thickness of the soot surface is 500 μm, as verified by the SEM measurements. This simple method is suitable for reproducible synthesis of uniform layers of carbon nanoparticles, the layer thickness depending on the deposition time [19]. The combustible material of candle wax is for the most part paraffin ($C_nH_{2n+2}$) and the burning process proceeds upwards in gravitational environment if adequate amount of oxygen is available. If the process is interfered by cooling the tip of the flame, the candle starts smoking and due to the incomplete burning process soot particles consisting of cyclic, highly unsaturated, polycyclic aromatic structural elements (($C_3H)_n$) are formed [29]. The primary soot particles grow through agglomeration, dehydration, and coagulation to as much as a few million carbon atoms and are deposited to the surface of a sample cup positioned at the tip of the flame (Fig. 2a).

The Nextel paint (Velvet-Coating 811-21, Mankiewicz Gebr. & Co) was included in the comparison because of its high emissivity over the wavelength range investigated in this work [26]. A painted surface is also more robust compared to the soot and nanotube samples and sustains even a light touch. The paint was applied directly on the surface of a sample cup by a professional painter by spraying with the required instruments and technique instructed in the datasheet of the manufacturer (Fig. 1g-h). The thickness of the Nextel coating was 200 μm.

In summary, the SEM measurements confirm that the Vantablack surface is highly uniform (Fig. 1c) as expected [30-32]. Its nanotube structure can be clearly distinguished in Fig. 1d. The S-VIS surface (Fig. 1a) is much rougher, but the magnification (Fig. 1b) reveals its nanotube nature. Soot surface is very smooth (Fig. 1f), with a surface roughness similar to that of Vantablack. The painted Nextel surface (Fig. 1h) is the most sealed one of the samples investigated here, leading to a reduced effective surface area. This is a likely explanation of Nextel surface's modest photoacoustic conversion efficiency, as discussed in the following section.

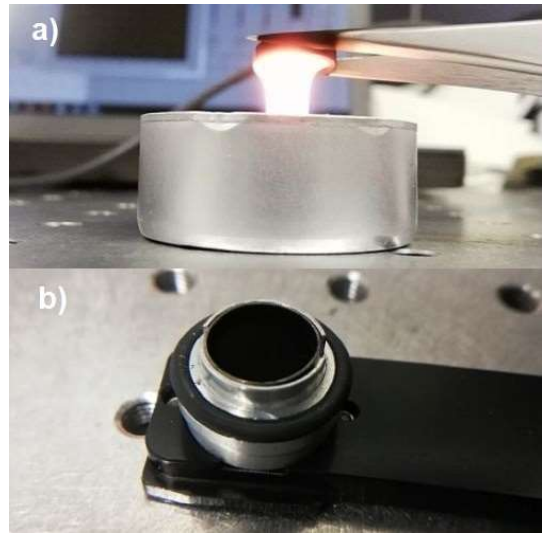

**Figure 2.** a) Sooting process, where a sample cup is held in the tip of a candle flame. b) A sample cup and holder of the PA301 photoacoustic detector.

## Measurements and results

The photoacoustic responsivities of the selected absorbers were compared over a broad spectral range, from the visible red (633 nm) to far infrared (25 μm). These comparisons were done using two complementary approaches. First, we measured the relative photoacoustic responsivities at several discrete wavelengths using monochromatic continuous-wave lasers. Second, similar measurements were carried out with an FTIR spectrometer that is equipped with a broadband incandescent light source. The FTIR measurements allowed us to extend the comparison to wavelengths inaccessible with lasers. In order to cancel out instrumental effects, all measurements were compared against the best-performing absorber, which in our case turned out to be the one based on candle soot.

### Laser measurements

The setup used for laser measurements is shown schematically in Fig 3. The laser power was modulated with a rotating-disk chopper before directing the laser beam to the absorber under study in normal-incidence configuration. The field of view seen by the detector was limited by irises to avoid any background radiation to be summed in the modulated laser radiation. Unless otherwise mentioned, the laser power levels were set to the same value (1.82 mW) by adjusting the laser drive current and/or neutral density filters. A very high signal-to-noise ratio (SNR) of over $10^4$ was obtained with this power level in the measurements reported here.

The photoacoustic signal recorded by the cantilever microphone was digitized, and the time-domain signal was subsequently Fourier transformed in real time to get the PA signal spectrum [15,33]. The actual signal proportional to the incident optical power was then obtained from the spectrum as the maximum value of the peak at the chopping frequency, using a recording time (Fourier time constant) of 1.57 s, unless otherwise mentioned. An example of a Fourier-transformed PA signal is shown in the inset of Fig. 3. The photoacoustic response depends on the chopping frequency, as illustrated in Fig. 4a for the candle soot-based absorber. (Similar plots for the other absorbers are presented in Fig. A1a of the Appendix). Note that this modulation-frequency dependence includes contributions from different parts of our PA301 photoacoustic detector, not just the absorber [15]. As an example, the figure clearly shows a mechanical resonance peak of the silicon cantilever. Although the cantilever microphone can be used with any modulation frequency within the range presented in Fig. 4, the best measurement SNR is typically obtained with frequencies below 100 Hz. At low frequencies below 20 Hz, the noise due to the external vibrations dominates. At higher frequencies the fundamental limit is set by the Brownian motion of the gas molecules when all external and electrical noise sources are eliminated [15].

Both the cantilever resonance frequency and the strength of the photoacoustic signal depend on the acoustic carrier gas. Due to its high thermal conductivity, helium is known to be one of the best choices in terms of signal maximization [34,35] and this was confirmed in our measurements. As an example, with the candle-soot absorber, helium gives up to 80 % larger signal than nitrogen, depending on the modulation frequency (Fig. 4b). The same tendency is observable also with the other absorbers investigated here – see Fig. A3 of the Appendix. The largest He/$N_2$ enhancement factor (of about 2.5) was obtained with the Nextel-painted surface.

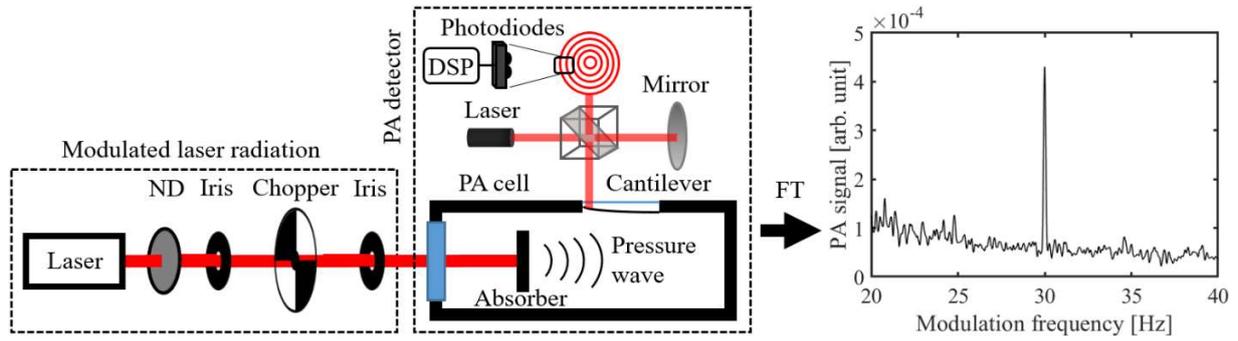

**Figure 3.** Simplified schematic of the photoacoustic laser power measurement setup (not all mirrors are shown). The chopped laser beam is aligned to the photoacoustic cell and the acoustic signal is recorded with the cantilever microphone by highly sensitive interferometric readout and processed with a Digital signal processor (DSP). The laser power level is adjusted with a neutral density filter (ND). The right-hand side of the figure shows an example of a PA spectrum, measured with 40 Hz chopping frequency, 6.26 s Fourier time constant and with an optical power of 50 nW (at 633 nm wavelength).

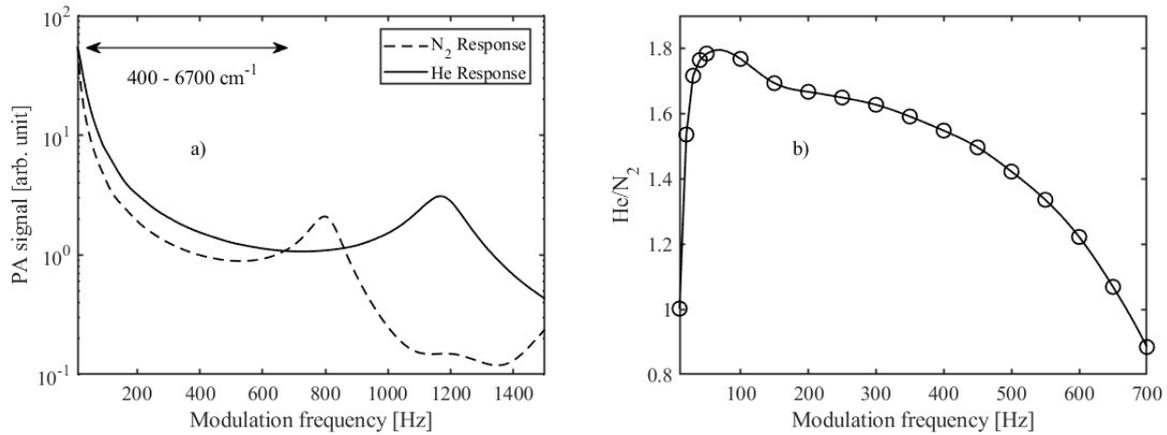

**Figure 4.** a) Photoacoustic response curves recorded with a 9.2 µm laser, candle-soot absorber and with two different carrier gases, He and $N_2$. The arrow indicates the wavenumber range covered in the complementary FTIR measurements (see next section and the Appendix). b) The ratio of these two curves in the frequency range of 10 to 700 Hz.

After characterizing the modulation-frequency dependence of the photoacoustic response, we compared the PA signals of the different absorber materials at six different wavelengths: 633 nm (He-Ne laser), 1064 nm (Yb-fiber laser), 1.63 µm (diode laser), 3.39 µm (He-Ne laser), 9.24 µm (Quantum Cascade Laser, QCL) and 14.85 µm (QCL). All lasers are continuous-wave lasers that produce highly monochromatic light. Other lasers are commercially available, but the 14.85 µm quantum cascade laser was custom-made for this work [36,37]. The result of this comparison is presented in Fig. 5a, which shows the absolute spectral responses of different absorbers with a chopping frequency of 40 Hz. Figure 5b shows the spectral responsivities divided by that of the candle-soot absorber, which gives clearly the strongest PA signal at all wavelengths. The error bars represent the estimated combined standard uncertainties, the dominant uncertainty sources being the reference power meter calibration and measurement repeatability (statistical measurement uncertainty).

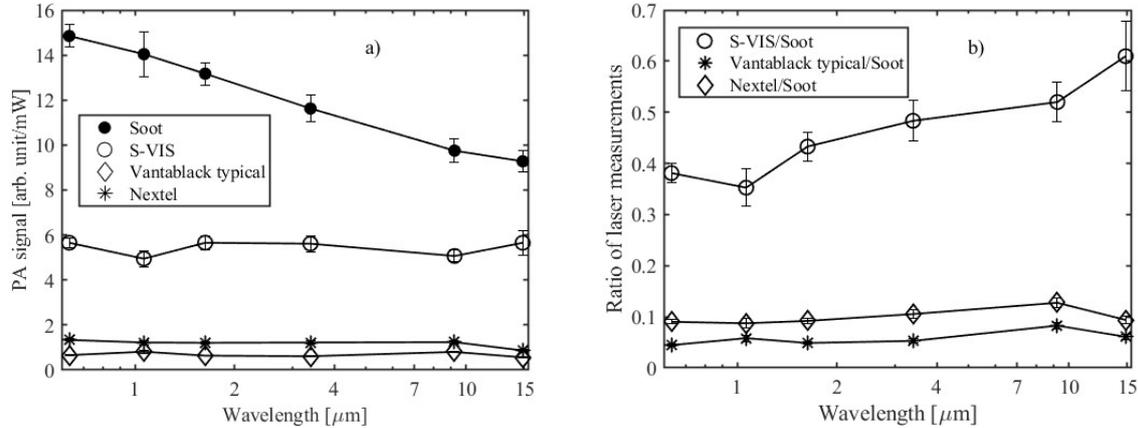

**Figure 5.** a) The photoacoustic signals of different absorbers normalized to 1 mW of optical power. b) The same spectral responsivities divided by that of the candle soot absorber, as measured with monochromatic lasers at six different wavelengths. The lines between the measured points are guides to the eye and do not present any physically meaningful fitting function. The chopping frequency was 40 Hz, and the acoustic carrier gas was helium.

**FTIR measurements**

The longest wavelength accessible in the laser measurements was 14.85 µm, which was achieved with our state-of-the-art quantum cascade laser technology. In order to extend the PA comparison to even longer infrared, we repeated the measurements with another setup (Fig. 6). The spectral range from 1.5 to 25 µm (6700 to 400 cm$^{-1}$) was continuously covered using a SiC thermal light source combined with an FTIR spectrometer (Bruker IRCube Matrix M series). The light emitted by the SiC was passed through the scanning Michelson interferometer of the FTIR instrument, thus producing a modulated output that was analyzed with the PA301 photoacoustic detector. As one of the mirrors of the Michelson interferometer is scanned at a constant speed $u$ the optical power component at wavelength $\lambda_o$ at the interferometer output is sinusoidally modulated at frequency $f = 2u/\lambda_o = (2u/c)\nu_o$, where $\nu_o$ is the optical frequency and $c$ is the speed of light. In our case of broadband light source, the interferogram (see the inset of Fig. 6) is the sum of the modulated signals of all wavelengths [38]. In other words, each optical frequency of the broadband light source is unambiguously mapped to a different acoustic frequency, and the Fourier transformed output of the PA301 detector (PA spectrum) is a down-converted replica of the original optical spectrum weighted by the spectral dependency of the PA detector, including the absorber. (The respective PA spectrum with optical wavelength on the horizontal axis can be recovered from the down-converted spectrum by multiplying the inverse of the acoustic-frequency axis by $2u$). We have chosen the FTIR mirror scanner speed ($u = 5.064 \times 10^{-4}$ m/s) such that the optical spectrum of the SiC light source is mapped to acoustic frequencies between 40.5 and 679 Hz, see the Appendix for details. This acoustic frequency range is well below the resonance frequency of the PA detector as indicated in Fig. 4a.

The photoacoustic FTIR spectra of different absorbers are presented in Fig. 7. Again, to cancel out the instrument function of the measurement setup (spectral variations of the light source, beam splitters, mirrors, etc.), we divided these spectra with that of the best absorber (candle soot) to get the relative PA responsivities as a function of optical wavelength. These ratios are shown in Fig. 8. Note that the individual photoacoustic FTIR spectra include contributions of the acoustic-frequency dependencies of the detector and the absorbers, because each optical frequency corresponds to a different acoustic frequency in the FTIR spectrum. In order to assist the interpretation of the FTIR measurements, we have plotted the modulation-frequency dependencies of different absorbers in the Appendix (Figs. A1a-b). Figure A1A indicates that the influence of modulation frequency on the S-VIS/Soot and Nextel/Soot ratios of Fig. 8 is small (except for the very long wavelengths), but the Vantablack/Soot ratios are strongly affected by this effect. The FTIR results were further validated with laser measurements, which are indicated by dots in Fig. 8. The laser measurements were carried out at modulation frequencies that correspond to the FTIR modulation frequencies of the respective optical wavelengths (see the Appendix Table A1).

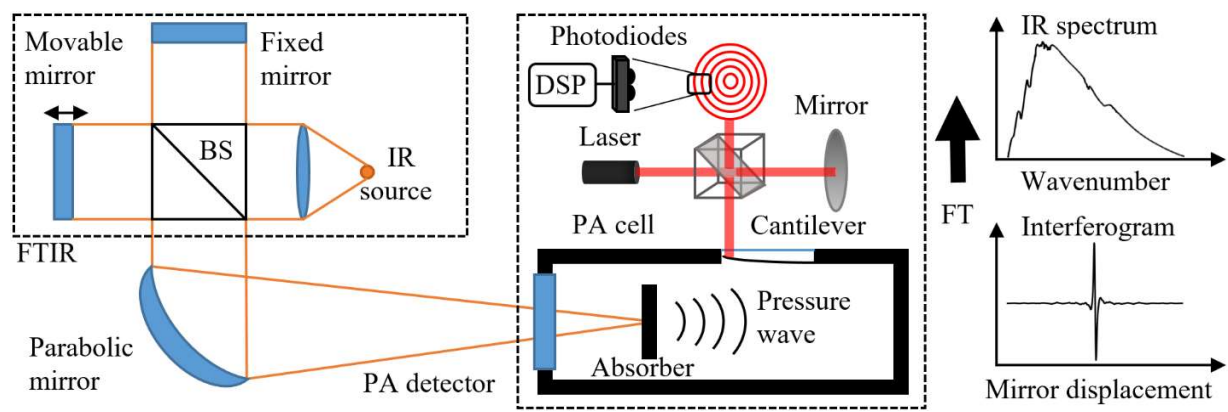

**Figure 6.** The principle of photoacoustic characterization of different absorbers using an FTIR spectrometer. Inside the FTIR, a broadband light emitted by the IR source (SiC) is modulated by the movable mirror of the interferometer. The collimated output is focused into the photoacoustic cell and the acoustic signal is recorded with the cantilever microphone by highly sensitive interferometric readout. Fourier transform (FT) of the interferogram gives the photoacoustic spectrum.

Despite the added complexity due to varying modulation frequency, the photoacoustic FTIR spectra give valuable complementary information to the laser measurements. As an example, Fig. 7 reveals that the spectral responsivities of other absorbers are smooth, but the Nextel coating loses its flat absorptivity above *c.a.* 2.8 μm and starts to act like a molecular absorber. The peak around 3 μm is a signature of the OH-group and the peaks between 3.3 and 3.6 μm and above 5 μm are mostly due to hydrocarbon molecular vibrations of the paint substances [16] (see Fig. A2e of the Appendix for a more detailed plot). It is also worth noting that the gradual improvement of the photoacoustic FTIR response of the vertically aligned CNT absorber (Vantablack) towards shorter wavelengths does not imply a real wavelength dependency, but rather reflects the improvement of Vantablack's responsivity with increasing modulation frequency (Fig. A1b).

The shaded areas around the curves in Fig. 8 describe the statistical uncertainties of the ratio measurements, as estimated from the standard deviations (1σ) of the FTIR measurements of each absorber. In order to calculate the standard deviations, the spectra were recorded 10 times for each sample. The raw data are presented in the Appendix, Fig. A2. The relative standard uncertainties are below 5 % for all the ratios over the entire measurement range – the increase of the uncertainty towards the edges of the spectral range is mostly due to the decreased spectral intensity of the light source at the detector (Fig. 7).

We also tested the reproducibility of sample preparation by producing and measuring multiple samples. With the candle-soot and S-VIS absorbers (5 of each), the maximum differences between the lowest and highest responsivities were 10 % and 15 %, respectively. Nextel surfaces were painted the same way and at the same time, and the differences between three different samples was less than 2 %. The highest variation was observed between different Vantablack samples, in which case "the best cut" gave 2.7 times higher PA signal than the worst one. This is illustrated in Fig. 8, which shows the FTIR curve of a typical Vantablack sample along with the best one. The reason for such a large variation between different Vantablack samples seems to be that it is practically impossible to cut the aluminum-foil substrate without bending it. For better reproducibility, the Vantablack absorbers should be grown directly on a substrate of the right size, such that unfolding of the vertically aligned CNT forest can be avoided. Similar sensitivity to substrate bending was not observed with other absorber materials.

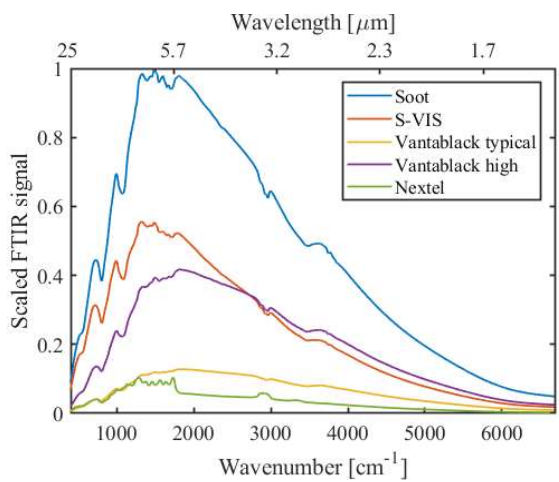

**Figure 7.** Photoacoustic FTIR spectra of different absorbers. All spectra are scaled by dividing them with the maximum signal of the soot sample. The spectral shape is mostly due to the SiC light source, whose emission spectrum closely follows Planck's law. The long-wavelength side of the spectrum is attenuated due to the increased losses of the FTIR's KBr beamsplitter at > 20 μm. The dips in the spectra are caused by absorbing molecules in the light path (mostly water vapor in the laboratory air). The spectral resolution of the FTIR instrument was set to 15 cm$^{-1}$, and the acoustic carrier gas used in the measurements was helium. For Vantablack, two curves are shown to exemplify the significant sample-to-sample variation, see text for details.

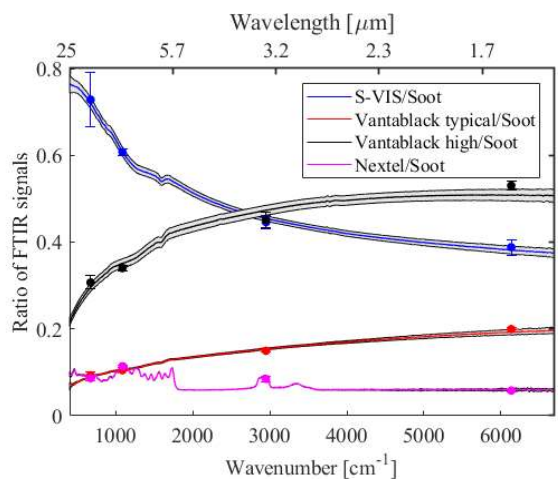

**Figure 8.** Photoacoustic signals of different absorbers divided by that of the candle soot absorber, as calculated from the PA FTIR spectra of Fig. 7. The shaded areas around the curves describe the statistical uncertainties of the FTIR measurements. Reference measurements done with lasers are indicated by dots and their statistical uncertainties by error bars.

## Conclusions

In conclusion, we have conducted an experimental comparison of photoacoustic responsivities of different highly absorptive materials. All of the tested absorbers – a candle soot absorber, two CNT absorbers and a Nextel-painted absorber – have nearly 100 % emissivities and look black when observed by naked eye. For example, the emissivity of Nextel-painted surface has been measured to be nearly constant 0.97 in the mid-infrared range, between 5 μm and 20 μm (500 cm$^{-1}$ to 2000 cm$^{-1}$) [26]. Despite the near-unity emissivities of the investigated absorber materials, the photoacoustic efficiencies vary significantly depending on the material, at least within the wavelength range (0.633 nm to 25 μm) and acoustic frequencies covered in our study. Self-made low-cost candle soot absorber was found to give the highest photoacoustic response within the entire spectral range, making it a good candidate for the future development of infrared power detectors. On the other hand, the PA responsivity of the candle soot absorber drops as a function of wavelength, while that of spray-coated CNT absorber (S-VIS) has a spectrally flat PA responsivity within our measurement uncertainties. The CNT absorbers appear promising especially in the longer infrared; their potential for THz photoacoustic detectors should be investigated. In all cases, the PA signal can be maximized by a proper choice of the acoustic carrier gas. For example, with the candle-soot absorber, helium provides up to 80 % enhancement of the PA signal compared to nitrogen, as measured with typical modulation frequencies between 40 Hz and 120 Hz.

## Acknowledgements

The work was funded by the Academy of Finland (Project numbers 314363 and 314364) and by the Academy of Finland Flagship Programme, Photonics Research and Innovation (PREIN), decision number: 320167. This work made use of Tampere Microscopy Center facilities at Tampere University.

# Appendix

**The effect of modulation frequency**

The FTIR ratios in Fig. 7 of the main article can be partly explained by the different modulation-frequency dependencies of different absorbers. These are exemplified below in Fig. A1, which shows the PA responses of the absorbers recorded at 14.85 μm wavelength and with modulation frequencies from 40 Hz to 650 Hz. The differences between different absorber types are clearly seen in Fig. A1b, where the modulation-frequency dependencies are compared to that of the candle soot absorber. In particular, the distinct differences in the modulation-frequency dependencies of Vantablack and candle soot are worth pointing out, as they explain the respective spectral shape in the FTIR measurement (Fig. 7 of the main article).

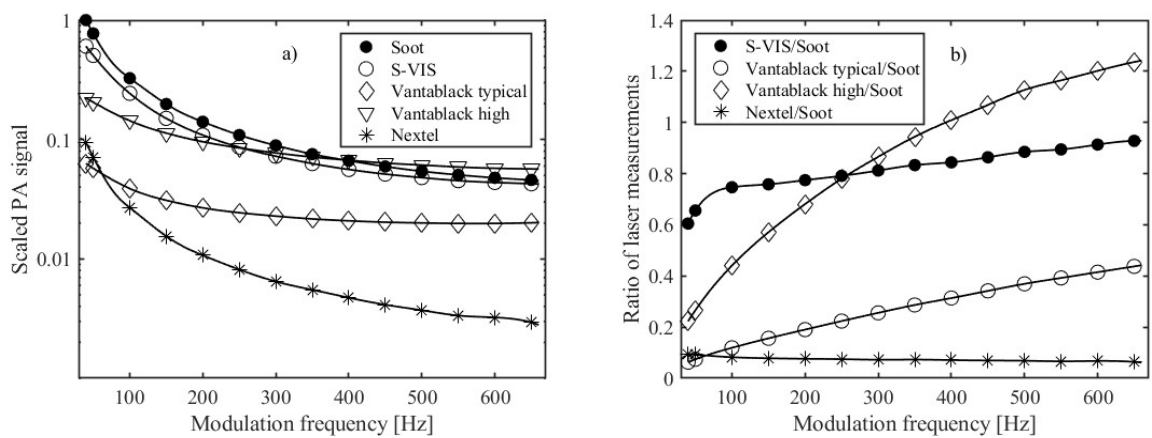

**Figure A1.** The photoacoustic response curves recorded with a 14.85 μm laser for all the absorbers in a). In b) ratios with respect to soot are presented. Carrier gas helium and modulation frequency range of 40 Hz to 650 Hz.

**FTIR measurements**

As explained in the main article, the FTIR spectrometer converts each optical wavelength to a different modulation (acoustic) frequency. In order to validate the FTIR measurements with laser measurements (Fig. 7 of the main article), we adjusted the laser chopping frequencies at each wavelength such that they equal to the respective FTIR modulation frequencies. These modulation frequencies $f$ [Hz] were calculated with equation $f = 2u\tilde{v}$, where $u = 0.05064$ cm/s is the FTIR mirror scanner speed and $\tilde{v}$ [cm$^{-1}$] is the wavenumber of light. (Note that this equation is equivalent to that shown in the main text for wavelength, since $\tilde{v}$ [cm$^{-1}$] = $10000/\lambda$ [μm], where $\lambda$ [μm] is the wavelength). The modulation frequencies calculated for the lasers used in the validation are presented in Table A1.

The FTIR average spectra in Fig. 6 were calculated from 10 separate measurements. These measurements were also used to calculate the standard deviations and the standard uncertainties plotted with shaded areas in Fig. 7. All the measured spectra are presented below in Fig. A2.

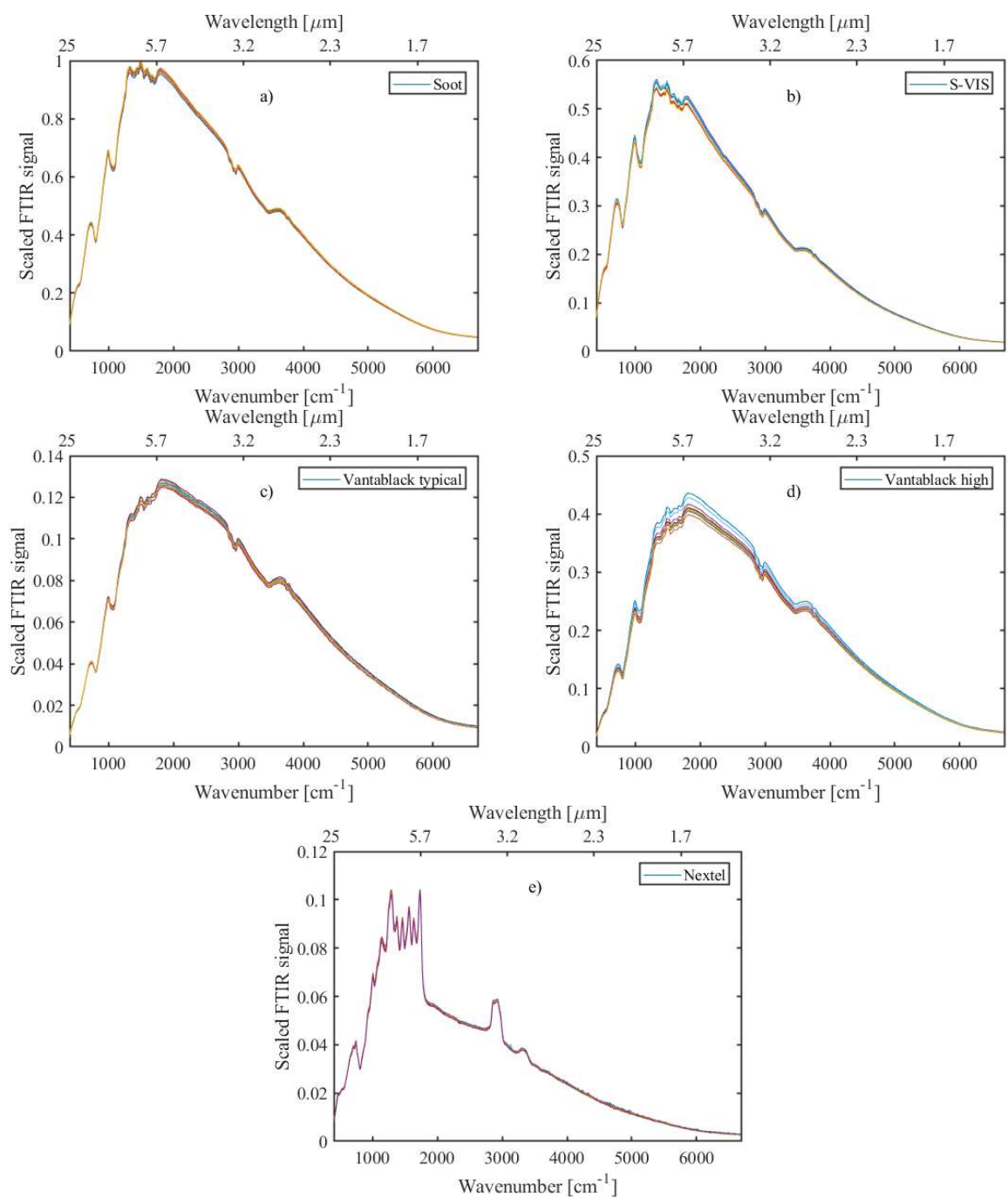

**Figure A2.** Photoacoustic FTIR spectra recorded with different absorbers. These data were used to calculate the average FTIR spectra and statistical uncertainties in Figs. 6 and 7 of the main article. Carrier gas was helium.

**Table A1.** Laser wavelengths along with the corresponding wavenumbers and modulation frequencies

| λ [μm] | ν [cm$^{-1}$] | f [Hz] |
|---|---|---|
| 1.63 | 6135 | 621 |
| 3.39 | 2950 | 299 |
| 9.24 | 1087 | 110 |
| 14.85 | 673 | 68.2 |

**The effect of acoustic carrier gas**

The carrier gas comparisons for all investigated absorbers are presented in Fig A3, as measured with the FTIR instrument. The wavenumber range of Fig. A3 corresponds to acoustic frequencies from 40.5 to 679 Hz (see Fig. 4a of the main article). The shapes of the FTIR curves depend on the modulation-frequency dependencies, as discussed above. (Also, see Fig. 4b of the main text; the similarity with the curve of Fig. A3 is apparent). The smooth Nextel-painted surface benefits the most from the exchange of the carrier gas. (The dips in the Nextel curve of Fig. A3 are caused by absorption peaks caused by water vapor. These absorption peaks are not visible with other absorbers due to the much higher measurement SNRs).

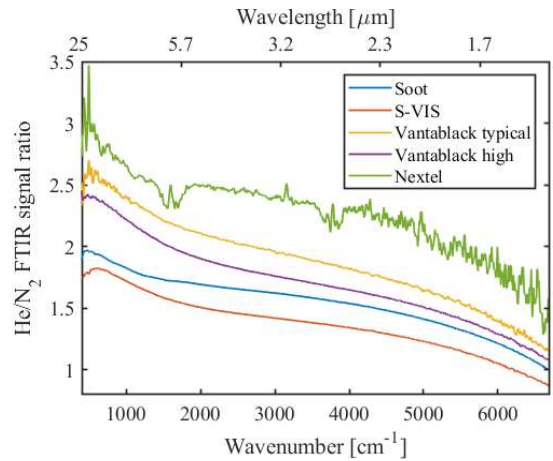

**Figure A3.** The ratios of photoacoustic FTIR signals measured with two different acoustic carrier gases, He and $N_2$.